\documentclass[10pt,twocolumn]{article}

\usepackage[a4paper,margin=2cm,columnsep=0.6cm]{geometry}
\usepackage{titlesec}
\usepackage{authblk}

\usepackage[T1]{fontenc}
\usepackage[utf8]{inputenc}
\usepackage{lmodern}
\usepackage{microtype}

\usepackage{amsmath,amssymb,amsfonts}

\usepackage{booktabs}
\usepackage{multirow}
\usepackage{graphicx}
\graphicspath{{fig/}}
\usepackage{float}
\usepackage{caption}
\usepackage{subcaption}
\usepackage{array}

\usepackage{algorithm}
\usepackage{algorithmic}

\usepackage{listings}
\usepackage[hidelinks,colorlinks=true,citecolor=blue!60!black,linkcolor=blue!60!black,urlcolor=blue!60!black]{hyperref}
\usepackage[numbers,sort&compress]{natbib}

\hypersetup{
  pdftitle={Cadence: Error-Bounded Lossy Compression of Demand Time Series with a Time-Series Foundation Model},
  pdfauthor={Roberto Tacconelli},
  pdfsubject={Error-Bounded Lossy Compression of Time Series},
  pdfkeywords={lossy compression, error-bounded compression, time-series foundation models, scientific data reduction, time-series databases},
  pdfusetitle=false,
}

\usepackage{xcolor}
\usepackage{enumitem}
\setlist{nosep,leftmargin=*}

\titleformat{\section}{\large\bfseries}{\thesection.}{0.5em}{}
\titleformat{\subsection}{\normalsize\bfseries}{\thesubsection}{0.5em}{}
\titleformat{\subsubsection}{\normalsize\itshape}{\thesubsubsection}{0.5em}{}

\newcommand{\sys}{\textsc{Cadence}}
\newcommand{\bpv}{\text{bpv}}
\newcommand{\tfm}{TimesFM-3}

\title{\textbf{Cadence: Error-Bounded Lossy Compression of\\Demand Time Series
with a Time-Series Foundation Model}}

\author[1]{Roberto Tacconelli}
\affil[1]{Independent Researcher\\
\texttt{tacconelli.rob@gmail.com}}

\date{}

\begin{document}
\maketitle
\thispagestyle{empty}

\begin{abstract}
We ask whether a pre-trained time-series foundation model reduces the bit
cost of numeric data, and find that the answer depends entirely on the coding
regime and the data domain. For \textbf{lossless} coding the answer is no, and
the reason is structural rather than an engineering deficit: code length
depends on the \emph{logarithm} of predictor accuracy, so the
1.51$\times$ accuracy advantage \tfm{}~\citep{timesfm3} holds over a 32-tap
linear predictor on real data buys $0.60$ bits out of $20.28$, or $2.9\%$.
Measured against classical predictors under an identical entropy coder,
\tfm{} yields a median gain of $\mathbf{+0.03\%}$ across 12 series---nil.
For \textbf{error-bounded lossy} coding the picture reverses on one specific
class of data. We introduce \sys{}, a closed-loop codec that guarantees
$|\hat{x}_t-x_t|\le\tau$ per sample, and evaluate it on two
\emph{uncontaminated} corpora postdating any plausible training cutoff:
49~US balancing-authority hourly demand series (EIA-930, 2026) and 50~MTA
subway station hourly ridership series (2026). Against the best of six
classical error-bounded predictors---every predictor coded by the same adaptive
arithmetic coder we implement---\sys{}
achieves a median gain of $\mathbf{+13.3\%}$ (147/147) on grid load and
$\mathbf{+28.3\%}$ (150/150) on ridership---\textbf{+21.4\% median over 297
series-tolerance pairs, winning all 297}. The same codec gains only $+6.4\%$
(21/24) on mixed operational telemetry and $+2.9\%$ on synthetic signals,
locating the effect in aggregate human-demand series rather than in numeric
data generally. We report five results that constrain how such systems must
be built and measured. (1)~The \emph{log$_2$ law} explains why forecasting
improvements do not transfer to lossless compression. (2)~\emph{Batch-size
invariance is unattainable}: no PyTorch configuration we tested makes
predictions bit-identical across batch sizes, so group size must be part of
the container format; we verify a bit-exact round trip under this constraint.
(3)~The \emph{context bootstrap} is a cost unique to neural codecs and
dominates short archives: end-to-end gains are $+6.8\%$ at six months of
hourly data, rising to $+15.1\%$ asymptotically, well below body-only figures.
(4)~The model's \emph{quantile head contributes nothing} beyond its median,
context length is worth $\sim$1 point, and a general-purpose entropy back end
is \emph{not neutral}---replacing xz/zstd with our coder gains $+9.7\%$ and
reverses one apparent finding. (5)~Against
\emph{downsampling}---what time-series databases actually deploy---\sys{}
delivers a worst-case error $28$--$56\times$ tighter at equal file size,
which we argue is the strongest practical case for the approach.
We test the domain claim by attempting to falsify it on SDRBench, where the
theory predicts failure and delivers it: $-0.8\%$ median, $0/27$, deepening to
$-41\%$ on a smooth simulation field.
\end{abstract}


\medskip
\noindent\textbf{Code and data:}
\url{https://github.com/robtacconelli/Cadence}

\section{Introduction}
\label{sec:intro}

Shannon~\citep{shannon1948} established that compression is prediction: a model
assigning high probability to the next symbol lets an encoder spend fewer bits
on it. Recent work has pushed this to large neural models, with Del\'etang et
al.~\citep{deletang2024} showing that a 70B language model compresses text
below classical context-mixing systems, and practical implementations
following~\citep{bellard2023,huang2024finezip,nacrith2026}.

Time-series foundation models---TimesFM~\citep{timesfm2023,timesfm3},
Chronos~\citep{chronos2024}, Moirai~\citep{moirai2024}---are the analogous
development for numeric sequences. They are pre-trained on trillions of time
points and forecast unseen series zero-shot. If compression is prediction, a
model that forecasts electricity demand better than a linear filter should
compress it better. This paper tests that inference and finds it substantially
false in the lossless case, and true only within a narrow domain in the lossy
case.

\begin{figure}[t]
\centering
\includegraphics[width=\columnwidth]{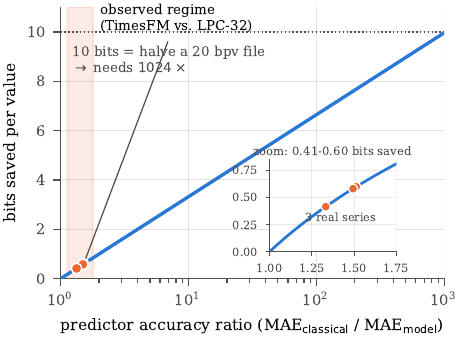}
\caption{Compression gain is logarithmic in forecasting gain
(Eq.~\ref{eq:log2}). Points are measured \tfm{}-versus-LPC-32 accuracy ratios
on three real series; the inset zooms the observed regime, where a
$1.33$--$1.51\times$ accuracy advantage yields $0.41$--$0.60$ bits. Halving a
20-bit-per-value file requires a $1024\times$ better predictor.}
\label{fig:log2}
\end{figure}

The negative result has a clean form. Code length under a well-matched
residual model is $\approx\log_2(\text{residual scale})+c$, so the bits saved
by a better predictor are logarithmic in the accuracy ratio:
\begin{equation}
\Delta \text{bits} \;=\; \log_2\!\left(\frac{\mathrm{MAE}_{\text{old}}}
{\mathrm{MAE}_{\text{new}}}\right).
\label{eq:log2}
\end{equation}
On real hourly pageview data, \tfm{} attains $\mathrm{MAE}=113{,}608$ against
$171{,}507$ for a 32-tap least-squares LPC---a genuine $1.51\times$
improvement---which Eq.~\ref{eq:log2} converts into $0.597$ bits out of
$20.28$, i.e.\ $2.9\%$. Halving a 20-bit-per-value file would require a
$\sim$$1000\times$ better forecaster. This single identity is sufficient to
rule out the lossless neural entropy coder, and with it per-file adaptation,
retrieval-augmented context, and lossless float coding, all of which route
their gain through a smaller residual. Figure~\ref{fig:log2} shows the
relationship together with the three measured operating points.

Error-bounded lossy coding escapes Eq.~\ref{eq:log2} at one point: when the
forecast lands inside the tolerance band the quantized residual is exactly
zero and the sample costs $\approx$0 bits. That is a discontinuity, not a
logarithm. We therefore build \sys{}, a closed-loop error-bounded codec, and
evaluate it against the predictors that state-of-the-art scientific
compressors actually use---the Lorenzo family and multilevel
interpolation of SZ3~\citep{sz3}, and ZFP~\citep{zfp}.

We make the following contributions.

\begin{enumerate}
\item \textbf{The log$_2$ law as a design constraint.} We formalize why
forecasting accuracy transfers only logarithmically into lossless compression
and verify it numerically, explaining a $+0.03\%$ median result that would
otherwise look like an implementation failure.

\item \textbf{An identical-coder evaluation protocol.} Comparing a neural
predictor against a classical one is only meaningful if both residuals pass
through the same entropy coder. We route every predictor through one adaptive
mixture-of-scales coder, and validate the harness on i.i.d.\ noise, where all
predictors read $12.003$ bits/value against a true entropy of $12.000$ and the
measured gain is exactly $+0.00\%$. An earlier version of our harness reported
a $+2.15\%$ ``gain'' on pure noise---an artifact of comparing a flexible
piecewise-linear density against a Laplace.

\item \textbf{Domain localization on uncontaminated data.} \tfm{} is
pre-trained on Wikipedia pageviews and Google Trends, so the obvious
benchmarks are contaminated. We evaluate on two corpora postdating any
plausible cutoff and show the gain is a property of \emph{aggregate
human-demand series} ($+21.4\%$, 297/297), not of numeric data
($-3.4\%$ synthetic, $+5.2\%$ mixed telemetry).

\item \textbf{A determinism constraint for neural codecs.} We show that model
predictions are \emph{not} bit-identical across batch sizes and that no
configuration we tested (TF32 disabled, SDPA disabled, deterministic
algorithms, forced MATH backend) repairs this. Encoder--decoder desynchronization
probability is $\sim$$8\times10^{-6}$ per sample, negligible per sample and
near-certain over a million. Group size must therefore be part of the container
format; we verify a bit-exact round trip under that rule.

\item \textbf{Context-bootstrap accounting.} A neural codec must transmit
$c$ samples of context that a Lorenzo predictor does not need. We price it,
show it dominates short archives, and report end-to-end rather than body-only
gains.

\item \textbf{Ablations.} Context length is worth $\sim$1 point, and the nine
quantiles---\tfm{}'s only probabilistic output---contribute $+0.3\%$ beyond
their median, so the codec can discard them.

\item \textbf{An adaptive arithmetic coder, and evidence that the back end is
not neutral.} We implement a binary range coder with context-modelled
binarization of quantization indices. It beats xz/zstd on real indices by
$+9.7\%$ (15/15), so general-purpose-back-end results understate neural and
classical predictors alike; more importantly it \emph{reverses} an apparent
finding (Section~\ref{sec:domains}) and re-explains a negative result
(Section~\ref{sec:ablation}) we had attributed to the wrong cause.

\item \textbf{Comparison against deployed practice.} Time-series databases
retain history by \emph{downsampling}, which has unbounded $L_\infty$ error.
At equal file size \sys{}'s guaranteed bound is $28$--$56\times$ tighter.

\item \textbf{Three negative results.} Cross-series conditioning through
covariates, foundation-model interpolation, and per-block hybrid switching all
fail; we report why, since two of the three fail for the same measurable
reason.
\end{enumerate}

\section{Related Work}
\label{sec:related}

\subsection{Error-Bounded Lossy Compression}
SZ~\citep{sz2016} and its successor SZ3~\citep{sz3} dominate error-bounded
scientific compression. Both quantize a prediction residual and entropy-code
the index. Critically, their predictors are \emph{small}: the Lorenzo
stencil~\citep{lorenzo2003} of order 1--3, linear regression, and in SZ3 the
hierarchical, anchor-based level-wise dynamic spline interpolation of Zhao et
al.~\citep{szinterp2021}, subsequently auto-tuned in
QoZ~\citep{qoz2024}. Our classical baseline family reimplements exactly these
predictors so that the comparison isolates prediction quality. ZFP~\citep{zfp} instead applies a block
transform, and FPZIP~\citep{fpzip} targets lossless float coding.
MGARD~\citep{mgard} provides multigrid error control. SDRBench~\citep{sdrbench}
is the standard corpus. This literature is overwhelmingly aimed at
multidimensional simulation fields; 1-D operational telemetry is not its
design target, a point that matters when interpreting our SZ3 comparison.

\subsection{Time-Series Compression in Databases}
Gorilla~\citep{gorilla2015} introduced XOR-based lossless float compression for
monitoring workloads and remains the basis of most production time-series
databases; Chimp~\citep{chimp2022} and Elf~\citep{elf2023} refine its XOR
coding, and Sprintz~\citep{sprintz2018} targets integer IoT streams with
delta coding and bit packing. Chiarot and Silvestri~\citep{tscompsurvey}
survey the area. For long-term retention these systems do not use error-bounded
codecs at all: they \emph{downsample} to coarse aggregates. This discards
extrema, which is precisely the information incident analysis needs, and it
provides no worst-case guarantee. We treat downsampling as the operative
baseline for the retention use case.

\subsection{Time-Series Foundation Models}
TimesFM~\citep{timesfm2023} introduced a decoder-only patched transformer for
zero-shot forecasting; \tfm{}~\citep{timesfm3} extends it to native
multivariate forecasting with 0.3B parameters trained on over a trillion time
points, using stacked variate attention and iterative reversible instance
normalization~\citep{revin2022}. It emits nine quantiles ($0.1$--$0.9$) per
horizon step and decodes a 64-step horizon non-autoregressively. Chronos
\citep{chronos2024} and Moirai~\citep{moirai2024} are contemporaneous.
Forecasting quality is well studied; the compression consequences are not,
which is the gap this paper addresses.

\subsection{Neural Compression}
NNCP~\citep{bellard2021} and CMIX~\citep{cmix2024} established that online
adaptation plus arithmetic coding yields state-of-the-art lossless ratios at
very low throughput. LLM-based text compressors~\citep{deletang2024,
bellard2023,huang2024finezip,nacrith2026} extend this to pre-trained models.
Our lossless finding is consistent with that literature and with
Eq.~\ref{eq:log2}: text gains are large because language models are
\emph{orders of magnitude} better than order-$k$ context models, whereas
\tfm{} is only $1.5\times$ better than a linear filter.

\section{Method}
\label{sec:method}

\subsection{Problem Setup}
Given an integer-valued series $x_{1:n}$ and an absolute tolerance $\tau$, an
error-bounded codec must emit a bitstream from which a decoder reconstructs
$\hat{x}_{1:n}$ with
\begin{equation}
|\hat{x}_t - x_t| \le \tau \qquad \forall t .
\label{eq:bound}
\end{equation}
We set $\tau = \rho\,\sigma(x)$ and sweep $\rho\in\{0.01,0.05,0.2\}$. Note that
$\rho$ maps to very different \emph{relative} errors by domain: $\rho=0.05$ is
$1.1\%$ on grid load but $4.5\%$ on station ridership, so comparisons across
domains should be made at matched relative error.

\subsection{Closed-Loop Quantization}
With quantization step $D=2\tau$ and predictor $P$,
\begin{align}
p_t &= P(\hat{x}_{t-c:t-1}), \qquad
k_t = \left\lceil \frac{x_t-p_t}{D} \right\rfloor, \\
\hat{x}_t &= p_t + k_t D ,
\end{align}
which satisfies Eq.~\ref{eq:bound} by construction. The predictor is fed its
own \emph{reconstruction}, never the original, so the decoder can reproduce
$p_t$ exactly. This closed loop is what makes the neural predictor's behaviour
under injected quantization noise relevant (Section~\ref{sec:noise}).

\subsection{Neural Predictor}
$P$ is \tfm{} with context $c$ and horizon 64, of which only the first step is
used, giving stride-1 operation. We take the median (quantile index 4) as the
point forecast. Section~\ref{sec:ablation} shows the other eight quantiles are
not worth transmitting, so we call the model with quantile output disabled.

\subsection{Group Size as a Format Parameter}
\label{sec:groups}
Because model outputs are not bit-identical across batch sizes
(Section~\ref{sec:determinism}), the format fixes a group size $G$; encoder and
decoder both run batches of exactly $G$. This is natural for a time-series
database, which compresses a block of $G$ metrics together in the manner of a
columnar block, but it does mean that decoding one series costs a full group.
We note that snapping predictions to a coarse grid does \emph{not} solve this:
it relocates the decision boundary rather than removing it, and a coarser grid
has more boundary per unit of drift.

\subsection{Context Bootstrap}
\label{sec:seed}
The model needs $c$ samples of history before it can predict, whereas a
Lorenzo-1 predictor needs one. \sys{} codes those $c$ samples \emph{lossily at
the same $\tau$} using the best of five side-information-free classical
predictors---Lorenzo orders 1--3, multilevel linear and cubic
interpolation---selecting per series and storing a one-byte identifier. LPC-32
is excluded because its least-squares coefficients would have to be
transmitted. The encoder then feeds the \emph{reconstructed} seed as model
context so that encoder and decoder share history from sample~0.

\subsection{Entropy Coding}
Quantization indices are coded by an adaptive binary range coder
(LZMA-style, 11-bit probabilities) with a CABAC-like binarization: a
context-coded zero flag, a bypass sign, a context-coded truncated-unary
magnitude prefix, and an Exp-Golomb bypass tail. Contexts derive from recent
magnitudes and are reproducible by the decoder, so nothing about them is
transmitted. Crucially, conditioning is expressed as \emph{contexts within one
stream} rather than as separate streams, so no partitioning can fragment the
coder---a failure mode that invalidated two of our earlier experiments. The
coder was validated by round-tripping Laplacian, sparse, heavy-tailed,
all-zero and uniform inputs. All reported figures are real bytes, and every
predictor---neural and classical---is coded by this same coder, so comparisons
remain comparisons of predictors.
Section~\ref{sec:ablation} reports idealized code lengths where they isolate a
modelling question, but never as headline results---in this study every
idealized figure proved optimistic relative to real bytes.

\begin{algorithm}[t]
\caption{\sys{} encode (one group of $G$ series)}
\label{alg:encode}
\begin{algorithmic}[1]
\REQUIRE series $x^{(1..G)}$, tolerances $\tau^{(1..G)}$, context $c$
\FOR{$g=1$ to $G$}
  \STATE $s^{(g)} \leftarrow \arg\min_{P\in\mathcal{S}} |\text{code}(P,x^{(g)}_{1:c},\tau^{(g)})|$
  \STATE $\hat{x}^{(g)}_{1:c} \leftarrow \text{reconstruct}(s^{(g)})$
  \COMMENT{lossy seed at the same $\tau$}
\ENDFOR
\FOR{$t=c+1$ to $n$}
  \STATE $p_{1:G} \leftarrow \tfm{}\big(\{\hat{x}^{(g)}_{t-c:t-1}\}_{g=1}^{G}\big)$
  \COMMENT{one batched pass of exactly $G$}
  \FOR{$g=1$ to $G$}
    \STATE $D \leftarrow 2\tau^{(g)}$;\;
           $k^{(g)}_t \leftarrow \lceil (x^{(g)}_t-p_g)/D \rfloor$
    \STATE $\hat{x}^{(g)}_t \leftarrow p_g + k^{(g)}_t D$
  \ENDFOR
\ENDFOR
\RETURN header $\|$ seeds $\|$ pack-and-compress$(k^{(g)})$
\end{algorithmic}
\end{algorithm}

\section{Experimental Setup}
\label{sec:setup}

\subsection{Hardware and Software}
All experiments run on a single NVIDIA RTX~5060 (8\,GB) with
PyTorch~2.12~\citep{pytorch2024} and \texttt{timesfm}~3.0.0. We note that the
\tfm{} model card supplies no separate citation for the third-generation
model: its BibTeX entry still points to the original decoder-only
paper~\citep{timesfm2023}, so we cite that work for the architecture lineage
and the model card and release note~\citep{timesfm3} for \tfm{}-specific
details. \tfm{} weights are 1.3\,GB and use 1.4--1.9\,GB of VRAM
at the batch sizes reported. SZ3 is built from source; ZFP is
\texttt{zfpy}~1.0.1. All figures are fp32; Section~\ref{sec:throughput}
discusses bf16.

\subsection{Baselines}
Our primary bar is the best of six classical error-bounded predictors, each run
closed-loop at the same $\tau$ with the identical entropy coder: Lorenzo orders
1--3, a 32-tap least-squares LPC, and multilevel linear and cubic
interpolation. Best-of-six is selected \emph{per row}, so the baseline is the
strongest classical result available rather than an average. We additionally
report the real SZ3 and ZFP binaries end-to-end, and downsampling with linear
reconstruction. We do not report \texttt{xz} as a lossy comparator: it is
lossless and the comparison would be meaningless.

\subsection{Corpora}
\textbf{Synthetic} (10 series $\times$ 20k samples) includes deliberate
controls: i.i.d.\ uniform noise, whose entropy any correct harness must
reproduce, and a random walk. \textbf{NAB}~\citep{nab2015} supplies 8 real
operational series (EC2 CPU/network/disk, RDS, autoscaling, NYC taxi, ambient
and machine temperature). \textbf{Grid} is EIA-930~\citep{eia930} hourly demand
for 49~US balancing authorities, January--June 2026. \textbf{Transit} is MTA
hourly ridership~\citep{mta2026} for the 50 busiest station complexes,
January--August 2026.

\textbf{Contamination control.} \tfm{} is trained on Wikipedia pageviews (to
November 2023) and Google Trends, so results on such series cannot support a
generalization claim. Grid and Transit both postdate any plausible cutoff and
carry the headline domain result. One balancing authority (\texttt{SEC}) is
excluded as corrupt: 3 of 4{,}343 samples carry sentinel values near
$-4.3\times10^{8}$ in a $\sim$300\,MW series, which inflates $\tau=\rho\sigma$
absurdly.

\subsection{Reproducibility}
All code, the experiment registry, and the JSON results behind every number in
this paper are available at \url{https://github.com/robtacconelli/Cadence}
under an MIT licence. Result files are committed, so every figure and table can
be regenerated without a GPU; scripts to rebuild each corpus from its primary
source are included. TimesFM-3 weights are not redistributed: they carry a
non-commercial licence, which this pipeline inherits.

\section{Results}
\label{sec:results}

\subsection{Lossless Coding Does Not Benefit}
\label{sec:lossless}

Table~\ref{tab:lossless} reports stride-1 lossless coding with every predictor
routed through one adaptive coder. The median gain over the best classical
predictor is $+0.03\%$ across 12 series, with 7/12 nominal wins of negligible
size.

\begin{table}[t]
\centering
\caption{Lossless coding, bits/value. \textsc{Classic} is the best of
\{LPC-32, LPC-256, seasonal-lag LS, 16-tap stack\}; all residuals pass through
the identical adaptive coder, so differences reflect prediction skill alone.}
\label{tab:lossless}
\footnotesize
\setlength{\tabcolsep}{3.5pt}
\begin{tabular}{@{}lrrrr@{}}
\toprule
\textbf{Series} & \textbf{xz} & \textbf{Classic} & \textbf{\tfm{}} & \textbf{Gain} \\
\midrule
ecg\_like       &  9.586 &  7.999 & \textbf{7.094} & $+11.3\%$ \\
sparse\_spiky   & 11.818 &  9.276 & \textbf{9.078} & $+2.1\%$ \\
wikihr\_en       & 23.858 & 19.758 & \textbf{19.288} & $+2.4\%$ \\
wikihr\_de       & 21.733 & 17.462 & \textbf{17.074} & $+2.2\%$ \\
regime\_switch  & 13.413 & 12.151 & \textbf{12.100} & $+0.4\%$ \\
poisson\_counts &  4.035 &  3.608 & \textbf{3.606} & $+0.1\%$ \\
seasonal\_metric & 15.611 & \textbf{12.399} & 12.417 & $-0.2\%$ \\
random\_walk    & 14.290 & \textbf{10.719} & 10.727 & $-0.1\%$ \\
\textit{iid\_noise}  & 12.736 & \textit{12.003} & \textit{12.003} & \textit{$+0.00\%$} \\
byte\_counter   & 17.379 & \textbf{8.099} &  8.219 & $-1.5\%$ \\
lorenz\_chaotic & 15.389 & \textbf{3.005} &  3.054 & $-1.6\%$ \\
\midrule
\textbf{Median}  & --- & --- & --- & $\mathbf{+0.03\%}$ \\
\bottomrule
\end{tabular}
\end{table}

The \texttt{iid\_noise} row validates the harness: true entropy is
$\log_2 4096 = 12.000$, every predictor reads $12.003$, and the gain is exactly
zero. This row is why we trust the rest of the table. It is also how we caught
an earlier harness defect that reported $+2.15\%$ on pure noise, which arose
from giving \tfm{} a flexible piecewise-linear density while the baseline was
locked to a Laplace---a measurement of density family, not of skill.

Eq.~\ref{eq:log2} accounts for the outcome. On \texttt{wikihr\_en}, \tfm{}
achieves $\mathrm{MAE}=113{,}608$ versus $171{,}507$ for LPC-32, a $1.513\times$
improvement worth $\log_2 1.513 = 0.597$ bits of a $20.28$-bit budget.

\subsection{Error-Bounded Lossy Coding, by Domain}
\label{sec:domains}

Table~\ref{tab:domains} summarizes all lossy evaluations under real-byte
accounting. The effect is sharply localized.

\begin{table}[t]
\centering
\caption{Median gain over best-of-six classical predictors, real bytes. Grid
and Transit postdate any plausible training cutoff.}
\label{tab:domains}
\small
\setlength{\tabcolsep}{4pt}
\begin{tabular}{@{}lrrr@{}}
\toprule
\textbf{Corpus} & \textbf{Median} & \textbf{Wins} & \textbf{Contam.} \\
\midrule
SDRBench (sci.)    &  $-0.8\%$ &  0/27 & n/a \\
Synthetic          &  $+2.9\%$ &  7/12 & n/a \\
NAB operational    &  $+6.4\%$ & 21/24 & no \\
Grid load (2026)   & $\mathbf{+13.3\%}$ & \textbf{147/147} & no \\
Transit (2026)     & $\mathbf{+28.3\%}$ & \textbf{150/150} & no \\
\midrule
\textbf{Demand combined} & $\mathbf{+21.4\%}$ & \textbf{297/297} & no \\
\bottomrule
\end{tabular}
\end{table}

\begin{figure}[t]
\centering
\includegraphics[width=\columnwidth]{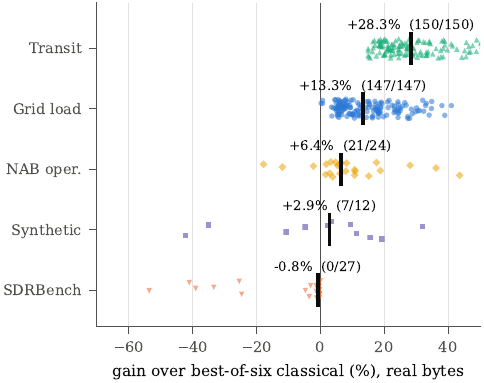}
\caption{Per-series-tolerance gain by corpus, real bytes. Each point is one
(series, $\rho$) pair; the bar marks the median. The effect is localized in
aggregate human-demand series, not in numeric data generally: it spans
$-0.8\%$ on scientific simulation output to $+28.3\%$ on ridership.}
\label{fig:domains}
\end{figure}

Figure~\ref{fig:domains} shows the full distribution behind
Table~\ref{tab:domains}: the two demand corpora separate cleanly from the
others, and their spread sits almost entirely above zero rather than being
carried by a tail.

Tables~\ref{tab:grid} and \ref{tab:transit} give the tolerance breakdown. All 297 series-tolerance pairs
gain---a clean sweep in both domains---and gains grow with tolerance in
\emph{both}, which is what the mechanism predicts: the advantage arises from
the discontinuity at zero residual, so it should compound as the band widens.

We initially reported the opposite for grid load, with gains apparently
\emph{shrinking} in $\rho$ ($+21.3\%$, $+14.0\%$, $+13.9\%$). That was an
artifact of the general-purpose back end: at loose tolerance a simple predictor
emits long runs of zeros, which LZMA compresses extremely well, flattering the
classical baseline exactly where \sys{} should pull ahead. With the arithmetic
coder the trend inverts. We report this because the retracted version is the
more publishable-looking result, and because it shows that delegating residual
coding to a generic compressor can manufacture a qualitative finding.

\begin{table}[t]
\centering
\caption{Grid load, 49 balancing authorities (EIA-930, 2026).}
\label{tab:grid}
\small
\begin{tabular}{@{}lrrrr@{}}
\toprule
$\boldsymbol{\rho}$ & \textbf{Median} & \textbf{Wins} & \textbf{Classic} & \textbf{\sys{}} \\
\midrule
0.01 & $+6.4\%$  & 49/49 & 4.356 & 4.009 \\
0.05 & $+13.3\%$ & 49/49 & 2.398 & 2.056 \\
0.20 & $+21.2\%$ & 49/49 & 1.265 & 0.980 \\
\midrule
\textbf{All} & $\mathbf{+13.3\%}$ & \textbf{147/147} & & \\
\bottomrule
\end{tabular}
\end{table}

\begin{table}[t]
\centering
\caption{Subway ridership, 50 station complexes (MTA, 2026).}
\label{tab:transit}
\small
\begin{tabular}{@{}lrrrr@{}}
\toprule
$\boldsymbol{\rho}$ & \textbf{Median} & \textbf{Wins} & \textbf{Classic} & \textbf{\sys{}} \\
\midrule
0.01 & $+19.9\%$ & 50/50 & 5.942 & 4.734 \\
0.05 & $+27.7\%$ & 50/50 & 3.324 & 2.374 \\
0.20 & $+51.2\%$ & 50/50 & 1.834 & 0.926 \\
\midrule
\textbf{All} & $\mathbf{+28.3\%}$ & \textbf{150/150} & & \\
\bottomrule
\end{tabular}
\end{table}

\begin{figure*}[t]
\centering
\includegraphics[width=\textwidth]{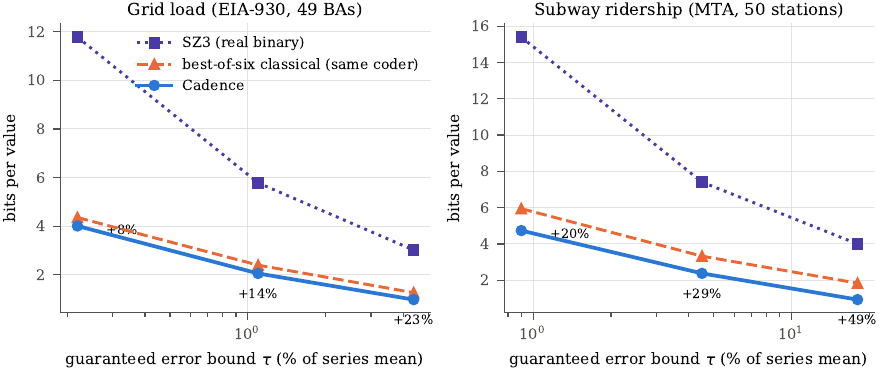}
\caption{Rate--distortion on the two uncontaminated demand corpora, plotted
against the \emph{guaranteed} error bound as a percentage of series mean.
Medians over 49 balancing authorities (left) and 50 station complexes (right);
annotations give \sys{}'s gain over best-of-six classical. SZ3 is the real
binary and is included for calibration, not as the primary bar
(Section~\ref{sec:domains}).}
\label{fig:rd}
\end{figure*}

Figure~\ref{fig:rd} plots the resulting rate--distortion curves against
guaranteed error rather than against $\rho$, which is the comparable axis
across domains.

Against the real SZ3 binary the median gain is $+36.3\%$, but this should be
read carefully: our own classical predictors also beat SZ3 on these data. The
honest reading is that SZ3, designed for multidimensional simulation fields, is
not the right tool for 1-D operational telemetry. Best-of-six at $+13.3\%$ is
the defensible bar. We also note that SZ3 carries roughly 500\,B of container
overhead, which dominates below $N\approx4$k; an earlier version of this
comparison at $N=2048$ overstated our advantage by a wide margin.

\subsection{SDRBench: A Falsification Test}
\label{sec:sdrbench}

Section~\ref{sec:domains} claims the effect belongs to demand series rather
than to numeric data. That claim predicts \sys{} should \emph{lose} on
scientific simulation output, where smooth fields make a local or interpolating
predictor near-optimal. We ran SDRBench~\citep{sdrbench} to try to falsify it:
six EXAALT molecular-dynamics trajectories and three Hurricane ISABEL
scanlines, evaluated 1-D against 1-D so that the comparison between predictors
remains fair.

\begin{table}[t]
\centering
\caption{SDRBench, bits/value, real bytes. The predicted loss is confirmed;
on the smooth field it deepens as the tolerance widens---the mirror image of
Table~\ref{tab:transit}.}
\label{tab:sdrbench}
\footnotesize
\setlength{\tabcolsep}{3.5pt}
\begin{tabular}{@{}lrrr@{}}
\toprule
\textbf{Subset} & \textbf{Median} & \textbf{Wins} & \textbf{vs.\ $\rho$} \\
\midrule
EXAALT (MD traj.) & $-0.6\%$ & 0/18 & flat \\
Hurricane (smooth) & $-25.4\%$ & 0/9 & deepens \\
\midrule
\textbf{All SDRBench} & $\mathbf{-0.8\%}$ & \textbf{0/27} & \\
\bottomrule
\end{tabular}
\end{table}

Table~\ref{tab:sdrbench} shows the prediction holds: not one of the 27
field-tolerance pairs gains. EXAALT trajectories, which are noisy and
effectively 1-D, are close to break-even ($-0.6\%$ median). Hurricane scanlines
lose heavily, and the loss \emph{grows} with tolerance ($-3.5\%$, $-25.4\%$,
$-41.0\%$ at $\rho=0.01,0.05,0.2$)---precisely inverted from ridership, where
gains grow with tolerance. Set against $+21.4\%$ on demand series, this makes
the domain characterization a tested boundary rather than an observation.

We note two caveats. Our codec is 1-D and cannot exploit the multidimensional
structure SZ3 is built for, so these numbers say nothing about SZ3 in its
native mode. And within this 1-D setting both \sys{} and our classical family
beat SZ3 by a wide margin, which again indicates that SZ3 in 1-D is being used
outside its design envelope.

\subsection{Predictor Behaviour Under Feedback Noise}
\label{sec:noise}

Because the closed loop feeds each predictor its own reconstruction, injected
quantization noise propagates. We measure the gain
$G = \mathrm{std}(P(x{+}\epsilon)-P(x))/\mathrm{std}(\epsilon)$.
Analytically, Lorenzo-1 has $G=1$, Lorenzo-2 $\sqrt5\approx2.24$, Lorenzo-3
$\sqrt{19}\approx4.36$, midpoint-linear interpolation $\sqrt{0.5}\approx0.707$
and 4-point cubic $0.80$. Measured, \tfm{} has $G\in[0.27,2.12]$ and LPC-32 up
to $5.44$.

A simple model, $\text{err}_{\text{tot}}^2 = \text{err}_{\text{clean}}^2 +
(G\,\sigma_\epsilon)^2$, predicts the win/loss sign in 15 of 18 cases,
including a full reversal on the Lorenz system where \tfm{} is $26.7\times$
\emph{less} accurate on clean data yet wins at large $\tau$. We stress the
conclusion this does \emph{not} support: \tfm{} is not contractive, whereas
SZ3's interpolation is contractive by construction. Noise robustness is
therefore not an unexploited gap---the field already exploits it, and better.
Against multilevel cubic interpolation \tfm{} loses on Lorenz at every
tolerance.

\subsection{End-to-End Codec and the Context Bootstrap}
\label{sec:endtoend}

\sys{} round-trips bit-exactly at $G=8$ and $G=16$: decoder reconstructions
match encoder reconstructions by SHA-256 and Eq.~\ref{eq:bound} holds on every
series. Table~\ref{tab:e2e} decomposes the container.

\begin{table}[t]
\centering
\caption{End-to-end container, $G=16$, $n=4300$, $\rho=0.05$, grid load.
Classical end-to-end on the same series, same coder, is $2.244~\bpv{}$.}
\label{tab:e2e}
\small
\begin{tabular}{@{}lrr@{}}
\toprule
\textbf{Component} & \textbf{\bpv{}} & \textbf{Share} \\
\midrule
Seed ($c=512$, coded lossily) & 0.412 & 20\% \\
Body (3{,}788 samples)        & 1.906 & 80\% \\
\midrule
\textbf{Total}                & \textbf{2.111} & \textbf{15.2$\times$} \\
\bottomrule
\end{tabular}
\end{table}

The bootstrap is a cost the classical baseline does not pay, and it dominates
short archives. Table~\ref{tab:length} shows the consequence: body-only figures
of $+13.3\%$ become $+6.8\%$ on six months of hourly data, converging to
$+15.1\%$ asymptotically.

\begin{table}[t]
\centering
\caption{End-to-end gain by series length, $\rho=0.05$, grid load, at the best
context for each length.}
\label{tab:length}
\small
\begin{tabular}{@{}lrrr@{}}
\toprule
\textbf{Length} & \textbf{Span} & \textbf{Best $c$} & \textbf{Gain} \\
\midrule
4{,}300   & 6 months & 512 & $+6.8\%$ \\
8{,}760   & 1 year   & 512 & $+11.0\%$ \\
17{,}520  & 2 years  & 512 & $+13.0\%$ \\
43{,}800  & 5 years  & 512 & $+14.2\%$ \\
\bottomrule
\end{tabular}
\end{table}

\subsection{Determinism}
\label{sec:determinism}

Table~\ref{tab:determinism} reports whether predictions are bit-identical
across call configurations. Repetition and batch \emph{reordering} are safe;
batch \emph{size} is not, and no configuration we tried repairs it.

\begin{table}[t]
\centering
\caption{Bit-identity of \tfm{} predictions, and attempts to force batch-size
invariance. Differences in MW on grid load.}
\label{tab:determinism}
\small
\begin{tabular}{@{}lcr@{}}
\toprule
\textbf{Condition} & \textbf{Identical} & \textbf{Max diff} \\
\midrule
Same batch, repeated $\times3$ & yes & $0$ \\
Batch reordered                & yes & $0$ \\
Batch 1 vs.\ batch 8           & \textbf{no} & $3.9\times10^{-3}$ \\
Batch 8 vs.\ batch 9           & \textbf{no} & $7.8\times10^{-3}$ \\
\midrule
\;+ TF32 disabled              & no & $7.8\times10^{-3}$ \\
\;+ \texttt{use\_sdpa=False}   & no & $7.8\times10^{-3}$ \\
\;+ deterministic algorithms   & no & $7.8\times10^{-3}$ \\
\;+ forced MATH backend        & no & $2.0\times10^{-3}$ \\
\bottomrule
\end{tabular}
\end{table}

With $D=985$\,MW at $\rho=0.05$ on MISO and a maximum drift of
$7.8\times10^{-3}$\,MW, the probability that a sample sits close enough to a
bin edge to flip is $\approx8\times10^{-6}$. That is negligible per sample and
near-certain across a million, and a single desynchronization destroys every
sample after it. Hence Section~\ref{sec:groups}.

\textbf{Cross-device portability.} We cannot test two machines, but we can test
something strictly harder on one: the same model and inputs executed on GPU
versus CPU. They are \emph{not} bit-identical (max $3.9\times10^{-3}$\,MW,
3/8 series). At $\rho=0.05$ that drift is $7.0\times10^{-6}$ of a quantization
step, giving an expected $\approx$$1.44\times10^{5}$ samples before the first
desynchronization: a six-month hourly series usually survives, a five-year
archive fails with probability $\approx$$26\%$, and beyond $\approx$$144$k
samples failure is effectively certain. The container must therefore record the
execution device as well as the group size. The obvious remedy---fp64---is
unavailable without patching the inference library, which pins its tensors to
fp32.

\subsection{Comparison with Deployed Retention Practice}
\label{sec:downsample}

At matched file size we compare against downsampling with linear
reconstruction, the mechanism production databases actually use.
Downsampling's worst-case error relative to \sys{}'s guaranteed bound is
$28.0\times$ (grid), $36.7\times$ (NAB) and $56.1\times$ (transit) at the
median. Downsampling does not claim an $L_\infty$ bound, so the fair statement
is narrow: wherever worst-case fidelity matters---incident forensics, anomaly
detection, compliance retention---downsampling is the wrong instrument and an
error-bounded codec is strictly better at the same cost.

\subsection{Throughput}
\label{sec:throughput}

Table~\ref{tab:throughput} shows that our initial throughput figure was an
artifact of batching, not a property of the model. Every (series, tolerance)
pair is an independent closed loop, so they advance in lockstep on one forward
pass.

\begin{table}[t]
\centering
\caption{Throughput. The model is compute-bound and saturates near batch
128--256.}
\label{tab:throughput}
\small
\begin{tabular}{@{}lrrr@{}}
\toprule
\textbf{Batch} & \textbf{Ctx} & \textbf{Prec.} & \textbf{Values/s} \\
\midrule
6   & 1024 & fp32 & 45 \\
64  & 1024 & fp32 & 113 \\
256 & 1024 & fp32 & 123 \\
256 &  512 & fp32 & \textbf{224} \\
256 &  512 & bf16 & \textbf{442} \\
\bottomrule
\end{tabular}
\end{table}

We report fp32 throughout. bf16 roughly doubles throughput but perturbs predictions by
$288$\,MW against a model error of $565$\,MW---$51\%$---which would inflate
residuals by roughly $12\%$. Encoder and decoder would still agree, since both
follow the same path, so bf16 remains viable for deployment; we simply decline
to conflate a speed choice with a compression result.

\subsection{Ablation Study}
\label{sec:ablation}

\begin{table}[t]
\centering
\caption{Context-length ablation, grid load, 16 series, identical scored window.
Seed cost is priced with the best-of-five classical coder at the same $\tau$.}
\label{tab:ablation-ctx}
\footnotesize
\setlength{\tabcolsep}{4pt}
\begin{tabular}{@{}rrrrrr@{}}
\toprule
\textbf{Ctx} & \textbf{Body} & \textbf{$\Delta$ body} & \textbf{Seed}
& \textbf{$n{=}4300$} & \textbf{$n{=}43800$} \\
\midrule
64   & 2.183 & $-11.6\%$ & 6.367 & $-0.0\%$ & $+2.5\%$ \\
128  & 2.071 & $-5.9\%$  & 4.582 & $+4.4\%$ & $+7.4\%$ \\
256  & 2.017 & $-3.1\%$  & 3.611 & $+5.9\%$ & $+9.7\%$ \\
512  & 1.959 & $-0.2\%$  & 3.214 & $\mathbf{+6.0\%}$ & $\mathbf{+12.0\%}$ \\
1024 & 1.956 & ---       & 2.980 & $+2.0\%$ & $+11.8\%$ \\
\bottomrule
\end{tabular}
\end{table}

\begin{figure*}[t]
\centering
\includegraphics[width=\textwidth]{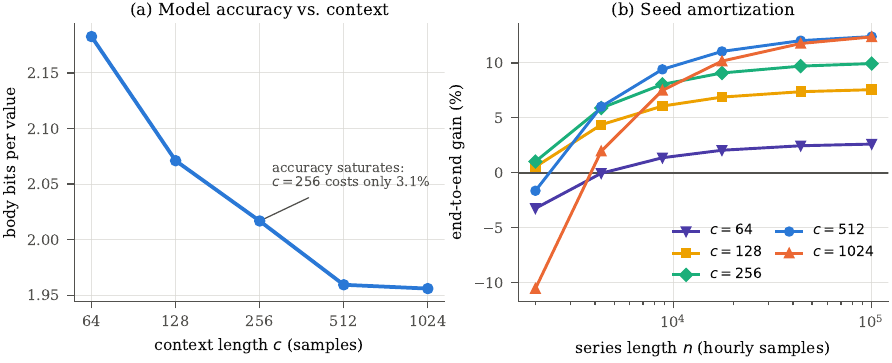}
\caption{Context ablation. (a)~Model accuracy saturates early: $c{=}256$ costs
only $1.6\%$ of body rate against $c{=}1024$ while quartering inference cost.
(b)~End-to-end gain against series length. Short contexts amortize their seed
quickly but code the body worse; long contexts are the reverse. The crossing
puts the optimum at $c{=}256$ for six months of hourly data and $c{=}512$
beyond one year.}
\label{fig:ablation}
\end{figure*}

\textbf{Context length} (Table~\ref{tab:ablation-ctx} and
Figure~\ref{fig:ablation}) is worth about two points at long archive lengths.
We had expected a large effect, reasoning that a shorter bootstrap costs
proportionally less; in fact body accuracy degrades about as fast as seed cost
falls, and $c=512$ is optimal at every length we tested. The practically useful
observation is different: accuracy saturates early, so $c=256$ costs only
$3.1\%$ against $c=1024$ while quartering inference cost. That is a systems result, not a compression one.

\textbf{The quantile head} contributes essentially nothing. Conditioning the
residual model on the predicted spread---the natural use of \tfm{}'s nine
quantiles---yields a median $+0.3\%$ (10/15 configurations) and is negative at
$\rho=0.2$. A flat adaptive coder already tracks the residual scale causally,
and predicted spread varies slowly enough to add no information. The codec can
therefore disable quantile output entirely.

We tested this three ways, and the diagnosis changed. Splitting the index
stream into eight spread buckets and compressing each separately reads
$-66.3\%$ (0/15), which measures fragmentation rather than modelling. Idealized
code lengths give $+0.3\%$. But expressing spread as \emph{contexts inside one
arithmetic-coded stream}---where fragmentation is impossible by
construction---still gives $\mathbf{-13.8\%}$ (0/15). The cause is therefore
not fragmentation but \emph{context dilution}: splitting the adaptive model
across eight contexts slows convergence more than the conditioning gains, and
predicted spread carries no information beyond the recent-magnitude context
already in use. Three independent implementations agree, and only the third
identifies the mechanism.

\subsection{Negative Results}
\label{sec:negative}

\textbf{Cross-series conditioning.} \tfm{} accepts covariates spanning context
and horizon. Conditioning \texttt{en.wikipedia} hourly views on
\texttt{de.wikipedia}---supplying the full trajectory of a correlated
series---moved code length from $19.045$ to $19.054~\bpv{}$, despite reducing
MAE from $113.6$k to $111.4$k.

\textbf{Foundation-model interpolation.} Since anchored interpolation beats
extrapolation on smooth signals, we coded a coarse grid, upsampled its
reconstruction, and supplied it as a past-and-future covariate. This failed in
every configuration tested ($K\in\{8,16\}$, stride $\in\{32,64\}$,
$\rho\in\{0.01,0.05\}$), from $-0.7\%$ to $-210\%$. Together with the previous
result this is two independent failures of the covariate pathway, and we
provisionally conclude it is not a usable side-information channel for coding.

\textbf{Hybrid prediction.} We resolved this with a design that needs \emph{no}
side information at all. Because the decoder holds the same reconstructed
history, it can recompute a causal classical prediction and both predictors'
error histories itself; a switch therefore costs zero bits. We ran three closed
loops---\tfm{} alone, a causal inverse-error \textsc{blend}, and
\textsc{leader} (use whichever predictor won the previous 256-sample
block)---each producing a single index stream, so fragmentation is impossible.
\textsc{leader} selects \tfm{} in $100\%$ of blocks and degenerates exactly to
\tfm{} alone; \textsc{blend} is \emph{worse} ($2.432$ vs.\ $2.221~\bpv{}$ at
$\rho=0.05$), since averaging with a weaker predictor hurts. The net gain is
$+0.0\%$: on this domain the model dominates the causal classical predictor
uniformly and there is nothing to hybridize. Our earlier $-4.0\%$ was entirely
fragmentation. \textsc{leader} is nonetheless worth shipping as free
insurance---it costs nothing and degenerates correctly off-domain, which
Section~\ref{sec:sdrbench} shows is a real operating regime.

\section{Discussion}
\label{sec:discussion}

\subsection{Why Demand Series}
The winning corpora share a generative structure: they are aggregate counts of
many independent human decisions, sampled at 30-minute to hourly resolution,
with strong daily and weekly periodicity. The losing corpora are per-machine physical and resource
metrics---\hspace{0pt}temperature, CPU load, disk throughput---\hspace{0pt}where a one- or
two-tap predictor is already near-optimal and
there is nothing for a pre-trained model to contribute. This is consistent with
Eq.~\ref{eq:log2}: gains require a large accuracy \emph{ratio}, and such ratios
only arise where the classical predictor is genuinely weak. SDRBench
(Section~\ref{sec:sdrbench}) marks the far end of that spectrum: on a smooth
simulation field the classical predictor is near-optimal, and \sys{} loses by
up to $41\%$.

We tried and failed to find a cheap statistic that predicts membership. Neither
the ratio of local to long-lag classical code length ($r=-0.37$) nor
daily-versus-lag-1 autocorrelation ($r=+0.44$, $n=49$) separates winners from
losers reliably. Practitioners should measure on a sample of their own data
rather than rely on a proxy.

\subsection{On Reporting Discipline}
Every idealized or projected number in this study came in high when
re-measured as real bytes end-to-end. Idealized code lengths overstated gains
by roughly a quarter; a projected $+9.9\%$ end-to-end figure measured $+2.9\%$;
a domain gain of $+15\%$ inferred from an idealized coder measured $-12.9\%$.
We recommend that work in this area report real bytes and treat idealized code
lengths strictly as upper bounds. A second discipline follows from
Section~\ref{sec:domains}: the entropy back end is part of the experimental
design. Delegating residual coding to xz or Zstandard understated every
predictor here by $\approx$$10\%$ and manufactured a qualitative tolerance
trend that reversed under a real coder.

\subsection{Limitations}

\begin{enumerate}
\item \textbf{Throughput.} At 224 values/s (fp32) compression is orders of
magnitude slower than classical codecs, which run at MB/s. \sys{} is an
archival codec. Decoding is strictly sequential and equally slow.

\item \textbf{Model overhead.} The 1.3\,GB model must be present at both
endpoints, and its weights carry a non-commercial licence. As with all neural
compressors we treat it as a shared standard; a self-contained archive would
need $\sim$54\,GB of payload to amortize it.

\item \textbf{Group-size coupling.} Section~\ref{sec:determinism} forces $G$
into the format, so decoding a single series costs a full group. This is
acceptable for columnar database blocks and awkward otherwise.

\item \textbf{Deliverable gain is modest.} End-to-end gains are $+6.8\%$ at six
months and $+14.2\%$ at five years. The larger body-only figures measure the
model's predictive advantage, not file size.

\item \textbf{Domain narrowness.} Two demand corpora support the result. It
does not extend to physical sensors, smooth simulation output, or
random-walk-like financial series, and we have no verified evidence for retail,
call-centre or utility-metering data.

\item \textbf{Bitstream portability.} Measured, not merely suspected: GPU and
CPU execution disagree, and the resulting desynchronization rate implies a
five-year hourly archive fails with probability $\approx$$26\%$
(Section~\ref{sec:determinism}). Portability requires integer or fixed-point
inference, which the current library does not support.
\end{enumerate}

\subsection{Future Work}
Integer or fixed-point inference would make the bitstream portable across
hardware and remove the group-size constraint, which is the main obstacle to
deployment. Distilling a small student to imitate the median forecast would
address throughput, though Eq.~\ref{eq:log2} caps what it can recover.
An asymmetric-numeral-systems back end~\citep{duda2009} would raise coding
throughput over our binary range coder without changing the model. Finally,
the downsampling comparison suggests the most useful framing is not
competition with SZ3 but replacement of lossy retention tiers in time-series
databases, which would benefit from evaluation inside a real database engine.

\section{Conclusion}
\label{sec:conclusion}

Forecasting accuracy and compression are related logarithmically, not linearly,
and that single fact determines where a time-series foundation model can help.
For lossless coding it cannot: a $1.5\times$ better forecaster buys $0.6$ bits
of a $20$-bit budget, and we measure a median gain of $+0.03\%$. For
error-bounded lossy coding on aggregate human-demand series it can: \sys{}
achieves $+21.4\%$ over the best of six classical predictors across 297
series-tolerance pairs on two uncontaminated 2026 corpora, winning all 297. The
deliverable end-to-end gain is smaller, $+6.8\%$ at six months rising to
$+15.1\%$ asymptotically, once the context bootstrap is paid for. The strongest practical
case is not against scientific compressors at all but against downsampling, the
mechanism time-series databases actually deploy, where \sys{} offers a
worst-case error $28$--$56\times$ tighter at equal size. Code, data pipeline
and the full experiment registry---including the eight claims we retracted under
better measurement---are at \url{https://github.com/robtacconelli/Cadence}.

\bibliographystyle{plainnat}

\end{document}